\documentclass[12pt]{article}
\usepackage{latexsym}
\usepackage{graphics}
\usepackage{color}
\begin{document}

\begin{titlepage}
\begin{flushright}
       {\bf UK/99-14}  \\
       Oct. 1999      \\
\end{flushright}
\begin{center}

{\bf {\LARGE  Parton Degrees of Freedom from the Path-Integral Formalism}}

\vspace{1.5cm}

{\bf  Keh-Fei Liu  \\
{\it National Center for Theoretical Sciences, P. O. Box 2-131, 
Hsinchu, Taiwan 300 \\
    and \\
 Dept. of Physics and Astronomy,  
  Univ. of Kentucky, Lexington, KY 40506}}
\end{center}

\begin{abstract}
We formulate the hadronic tensor $W_{\mu\nu}$ of deep inelastic scattering
in the path-integral formalism. It is shown that there are 3 gauge invariant 
and topologically distinct contributions. Besides the valence contribution,
there are two sources for the sea -- one in the connected
insertion and the other in the disconnected insertion. 
The operator product expansion is carried out in this formalism. The operator
rescaling and mixing reveal that the connected sea partons evolve
the same way as the valence, i.e. their evolution is decoupled from the
disconnected sea and the gluon distribution functions. We explore the 
phenomenological consequences of this classification in terms of the 
small x behavior, Gottfried sum rule violation, and the flavor dependence.
In particular, we point out that in the nucleon 
$\bar{u}$ and $\bar{d}$ partons have both the connected and disconnected sea
contributions; whereas, $\bar{s}$ parton has only the disconnected sea 
contribution. This difference between $\bar{u} + \bar{d}$ and $\bar{s}$, as 
far as we know, has not been taken into account 
in the fitting of parton distribution functions to experiments.


\bigskip
 
 
\end{abstract}

\vfill
\end{titlepage}

\section{Introduction}

In the last decade, the surprising results of a small
quark spin content (flavor-singlet $g_A^0$)~\cite{emc88}
 and the discovery that $\bar{u} \neq \bar{d}$ in the nucleon
~\cite{nmc92} from deep inelastic scattering have focused 
people's
attention on the interplay between the parton model
at high energies and the hadronic structure at low
energies. The connection is often made through the
operator product expansion which relates the sum rules
of parton distribution functions to the forward
matrix elements. The latter can be obtained from low energy
experiments.   
 
In the parton model, the dynamical quark degrees of freedom
are taken to be the valence and the sea. Whereas, in the hadronic
models, the degrees of freedom involve the valence and the meson cloud.
The classification of these dynamical degrees of freedom in
deep inelastic scattering has been made in the path integral
formalism~\cite{ld94,ldd99}. It is revealed that there are
two sources for the sea quark. One is in the connected insertion and
the other in the disconnected insertion. Their relations to the meson 
cloud in the hadronic models for hadrons near the rest frame and
chiral symmetry have been clarified and extensively explored
in the context of hadronic models in terms of 
the form factors, hadron masses, and
matrix elements, i.e. low-energy quantities which can be
calculated in the two- and three-point functions. 
In addition, it is shown that when
both the connected sea (referred to as cloud in Ref.\cite{ldd99}) and 
the disconnected sea quarks
are eliminated in a valence QCD theory, the valence quark picture with 
SU(6) symmetry emerges. In this paper, we shall derive the operator 
product expansion in the path-integral formalism and explore the 
phenomenological consequences of this classification of the parton degrees 
of freedom.
Section 2 is on the path-integral formalism of the hadronic tensor and the
classification of the parton degrees of freedom. Section 3 shows how to
carry out the operator product expansion in the path-integral formalism
and to derive parton evolution equations through operator rescaling and
mixing. Section 4 explores the phenomenological consequences.
We shall show that the small x behavior of the connected sea is
different from that of the disconnected sea and the violation of the
Gottfried sum rule, i.e. $\bar{u} \neq \bar{d}$, comes only from the
connected sea at the flavor $SU(2)$ limit. Finally, we emphasis that $\bar{u} +
\bar{d}$ in the nucleon has both the connected sea and the disconnected sea
contributions; whereas $\bar{s}$ has only the disconnected sea contribution.
This difference has not been parametrized in extracting the parton distribution
functions from the experiments. The conclusion is given in Section 5.

\section{Path-Integral Formalism}  \label{pif}

The deep inelastic scattering of a muon on a nucleon involves the hadronic
tensor which, being an inclusive reaction, involves all intermediate states
\begin{equation}   \label{w}
W_{\mu\nu}(q^2, \nu) = \frac{1}{2M_N} \sum_n  (2\pi)^3 
\delta^4 (p_n - p - q) \langle N|J_{\mu}(0)|n\rangle
\langle n|J_{\nu}(0) | N\rangle_{spin\,\, ave.}. 
\end{equation}
Since deep inelastic scattering measures the absorptive part of the  
Compton scattering, it is the imaginary part of the forward amplitude and
can be expressed as the current-current correlation function in the nucleon, 
i.e.
\begin{equation}  \label{wcc}
W_{\mu\nu}(q^2, \nu) = \frac{1}{\pi} Im T_{\mu\nu}(q^2, \nu)
= \frac{1}{2M_N}
\langle N| \int \frac{d^4x}{2\pi}  e^{i q \cdot x} J_{\mu}(x)
J_{\nu}(0) | N\rangle_{spin\,\, ave.}.
\end{equation} 

It has been shown~\cite{ld94,ldd99} that the hadronic tensor 
$W_{\mu\nu}(q^2, \nu)$ can be obtained from the Euclidean path-integral
formalism where the various parton dynamical degrees of freedom are
readily and explicitly revealed.
In this case, one considers the ratio of the four-point function 
\mbox{$\frac{2 E_p V}{2 M_N}\langle O_N(t) \int \frac{d^3x}{2\pi} 
e^{-i \vec{q}\cdot  \vec{x}} J_{\nu}(\vec{x},t_2) J_{\mu}(0,t_1)
O_N(0)\rangle$} and the two-point function
\mbox{$\langle O_N(t-(t_2-t_1)) O_N(0)\rangle$},
where $O_N(t)$ is an interpolation
field for the nucleon with momentum $p$ at Euclidean time $t$.
 
As both $t - t_2 >> 1/\Delta E_p$ and $t_1 >> 1/\Delta E_p$, where
$\Delta E_p$ is the energy gap between the nucleon energy $E_p$ and the next
excitation (i. e. the threshold of a nucleon and a pion in the $p$-wave),
the intermediate state contributions will be dominated by
the nucleon with the Euclidean propagator $e^{-E_p (t-(t_2 - t_1))}$.
Hence,
\begin{eqnarray}  \label{wmunu_tilde}
\widetilde{W}_{\mu\nu}(\vec{q}^{\,2},\tau) &=&
 \frac{\frac{2 E_p V}{2M_N}< O_N(t) \int \frac{d^3x}{2\pi} e^{-i \vec{q}\cdot
 \vec{x}}
 J_{\mu}(\vec{x},t_2)J_{\nu}(0,t_1) O_N(0)>}{<O_N(t- \tau) O_N(0)>} \,
 \begin{array}{|l} \\  \\  \footnotesize{t -t_2 >> 1/\Delta E_p} \\
 \footnotesize{t_1 >> 1/\Delta E_p} \end{array} \nonumber \\
   &=& \frac{\frac{f^2 2 E_p V}{2M_N} e^{-E_p(t-t_2)}<N| \int \frac{d^3x}{2\pi}
 e^{-i\vec{q}\cdot \vec{x}} J_{\mu}(\vec{x},t_2) J_{\nu}(0,t_1)|N>
e^{-E_pt_1}}{f^2 e^{-E_p(t-\tau)}} \nonumber \\
  &=&\frac{2 E_p}{2M_N} <N|\int \frac{d^3x}{2\pi} e^{-i\vec{q}\cdot \vec{x}}
J_{\mu}(\vec{x},t_2) J_{\nu}(0,t_1)|N>,
\end{eqnarray}
where $\tau = t_2 - t_1$ and $f$ is the transition matrix element
$\langle 0|O_N|N\rangle$, and V is the 3-volume. Inserting intermediate states, 
$\widetilde{W}_{\mu\nu}(\vec{q}^{\,2},\tau)$ becomes
\begin{equation}   \label{wtilde}
\widetilde{W}_{\mu\nu}(\vec{q}^{\,2},\tau)
= \frac{1}{2M_N} \sum_n (2\pi)^2 
\delta^3 (p_n - p + q)  \langle N|J_{\mu}(0)|n\rangle
\langle n|J_{\nu}(0) | N\rangle_{spin\,\, ave.} e^{- (E_n - E_p) \tau}.
\end{equation}
To go back to the 
delta function $\delta(E_n - E_p + \nu)$ in Eq. (\ref{w}), one needs
to carry out the inverse Laplace transform~\cite{wil93,ld94}
\begin{equation}  \label{wmunu} 
W_{\mu\nu}(q^2,\nu) = \frac{1}{i} \int_{c-i \infty}^{c+i \infty} d\tau
e^{\nu\tau} \widetilde{W}_{\mu\nu}(\vec{q}^{\,2}, \tau),
\end{equation} 
with $c > 0$. This is basically doing the anti-Wick rotation back to the 
Minkowski space. 
 
In the Euclidean path-integral formulation of 
$\widetilde{W}_{\mu\nu}(\vec{q}^{\,2}, \tau)$ in Eq. (\ref{wtilde}),
contributions to the four-point function can be classified
according to different topologies of the quark paths between
the source and the sink of the proton. They represent different
ways the fields in the currents $J_{\mu}$ and $J_{\nu}$ contract with
those in the nucleon interpolation operator $O_N$.
Fig. 1(a) and 1(b) represent connected insertions (C.I.) of the
currents.  Here the quark fields from the interpolators $O_N$ contract
with the currents such that the quark lines flow continuously from $t =
0$ to $t =t$. Fig. 1(c), on the other hand, represents a disconnected
insertion (D.I.) where the quark fields from $J_{\mu}$ and $J_{\nu}$
self-contract and are hence disconnected from the quark paths between
the proton source and sink. Here, ``disconnected'' refers only to the 
quark lines. Of course, quarks dive in the
background of the gauge field and all quark paths are ultimately
connected through the gluon field.

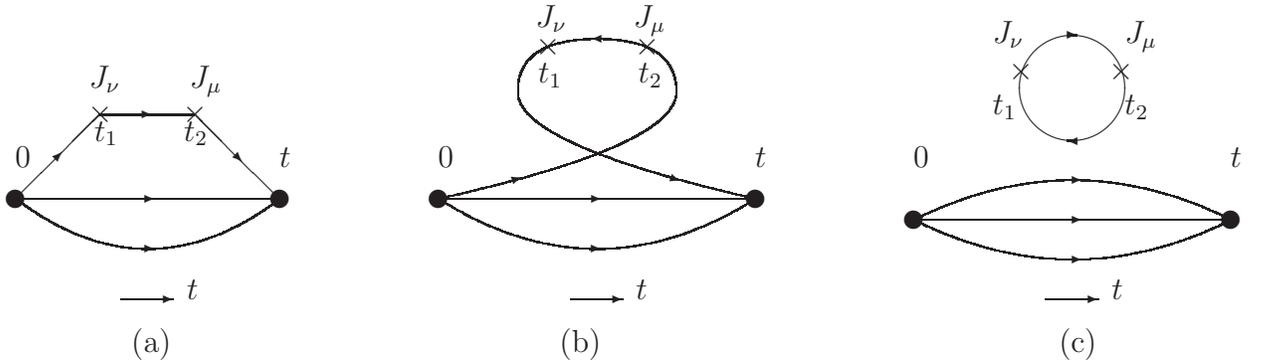
\begin{figure}[h]
\[
\hspace*{1.5in}\setlength{\unitlength}{0.01pt}
\begin{picture}(45000,20000)
\put(-10000, 8800){\circle*{700}}
\put(-10000, 8800){\line(1,1){3182}}
\put(-8410, 10391){\vector(1,1){200}}
\put(-7258,11700){{\bf $\times$}}
\put(-6818,11982){\line(1,0){3600}}
\put(-5018,11982){\vector(1,0){200}}
\put(-3658,11700){{\bf $\times$}}
\put(0000, 8800){\line(-1,1){3182}}
\put(-1591,10391){\vector(1,-1){200}}
\put(-10000, 8800){\line(1,0){10000}}
\put(-5000,  8800){\vector(1,0){300}}
\put(0000, 8800){\circle*{700}}
\qbezier(-10000, 8800)(-5000, 5000)(00, 8800)
\put(-5000, 6900){\vector(1,0){300}}
\put(-6000, 5000){\vector(1,0){2000}}
\put(-3500, 5000){$t$}
\put(-10000,10000){$0$}
\put(0,10000){$t$}
\put(-7000,11000){$t_1$}
\put(-3600,11000){$t_2$}
\put(-7200,13000){$J_\nu$}
\put(-3400,13000){$J_\mu$}
\put(-5600, 3000){(a)}
\put(6000, 8800){\circle*{700}}
\put(18000, 8800){\circle*{700}}
\qbezier(9000, 12850)(9000 ,14850)(12000,14850)
\qbezier(12000,14850)(15000,14850)(15000,12850)
\qbezier(6000 , 8800)(15000,10800)(15000,12850)
\qbezier(9000 ,12850)(9000 ,10850)(18000,08800)
\put(12000,14850){\vector(-1,0){200}}
\put(9000,  9550){\vector(3,1){200}}
\put(15000, 9550){\vector(3,-1){200}}
\put(9700, 14250){{\bf $\times$}}
\put(13450,14250){{\bf $\times$}}
\put(6000, 8800){\line(1,0){12000}}
\put(12000, 8800){\vector(1,0){200}}
\qbezier(6000,08800)(12000, 5000)(18000, 8800)
\put(12000, 6900){\vector(1,0){200}}
\put(13500, 5000){$t$}
\put(6000,10000){$0$}
\put(18000,10000){$t$}
\put(11000, 5000){\vector(1,0){2000}}
\put(9800 ,13200){$t_1$}
\put(13600,13200){$t_2$}
\put(9700 ,15300){$J_\nu$}
\put(13400,15300){$J_\mu$}
\put(10600, 3000){(b)}
\put(24000, 8000){\circle*{700}}
\put(24000, 8000){\line(1,0){12000}}
\put(30000, 8000){\vector(1,0){400}}
\put(36000, 8000){\circle*{700}}
\qbezier(24000, 8000)(30000, 5000)(36000, 8000)
\put(30000, 6500){\vector(1,0){400}}
\qbezier(24000, 8000)(30000,11000)(36000, 8000)
\put(30000, 9500){\vector(1,0){400}}
\put(30000,13000){\circle{4000}}
\put(27600,13300){{\bf $\times$}}
\put(31400,13300){{\bf $\times$}}
\put(30000,15000){\vector(1,0){200}}
\put(30000,11000){\vector(-1,0){200}}
\put(29000, 5000){\vector(1,0){2000}}
\put(31500, 5000){$t$}
\put(24000,10000){$0$}
\put(36000,10000){$t$}
\put(27000,12000){$t_1$}
\put(32000,12000){$t_2$}
\put(27000,14700){$J_\nu$}
\put(32000,14700){$J_\mu$}
\put(29500,3000){(c)}
%
\end{picture}
\]
\caption{Quark skeleton diagrams in the Euclidean path integral formalism
for evaluating $W_{\mu\nu}$ from the four-point function defined
in Eq. (\ref{wmunu_tilde}). These include the lowest twist contributions to
$W_{\mu\nu}$. (a) and (b) are connected insertions and (c) is a
disconnected insertion.}
\end{figure}


Fig. 1 represents the contributions of
the class of ``handbag" diagrams where the two currents are hooked
on the same quark line. These contain leading twist contributions in deep
inelastic scattering. Other contractions with the
two currents hooking on different quark lines involve only
higher twist operators and thus will be suppressed in the 
Bjorken limit~\cite{ldd99}. They are shown in Fig. 2. 
We will neglect these ``cat's ears" diagrams from now on.
We should stress that these diagrams in Figs. 1 and 2 are {\it not}
Feynman diagrams to repesent the forward Compton scattering amplitude
$T_{\mu\nu}(q^2, \nu)$ and should not be read as such. Rather, they
are path-integral diagrams needed to formualte 
$W_{\mu\nu}(q^2, \nu)$ which is the imaginary part of $T_{\mu\nu}(q^2, \nu)$
(see Eq. (\ref{wcc})) or its s-channel discontinuity.

\begin{figure}[h]
\[
\hspace*{1.5in}
\setlength{\unitlength}{0.01pt}
\begin{picture}(45000,20000)
\put(-10000,8800){\circle*{700}}
\put(-7258,10750){{\bf $\times$}}
\put(-2000, 8500){{\bf $\times$}}
\put(-10000,8800){\line(1,0){12000}}
\put(-4000, 8800){\vector(1,0){200}}
\put(-4000,  5900){\vector(1,0){200}}
\put(-4000, 11700){\vector(1,0){200}}
\put(2000,8800){\circle*{700}}
\qbezier(-10000,8800)(-4000, 3000)(2000, 8800)
\qbezier(-10000,8800)(-4000,14600)(2000, 8800)
\put(-5000, 5000){\vector(1,0){2000}}
\put(-3000, 5000){$t$}
\put(-10000,10000){$0$}
\put(2000,10000){$t$}
\put(-7000, 9600){$t_1$}
\put(-2000, 7300){$t_2$}
\put(-7200,12100){$J_\nu$}
\put(-2000, 9500){$J_\mu$}
\put(-4600, 3000){(a)}
\put(4500,8800){$+$}
\put(9000,8000){\circle*{700}}
\put(9000,8000){\line(1,0){12000}}
\put(15000,8000){\vector(1,0){200}}
\put(21000,8000){\circle*{700}}
\qbezier(9000,8000)(15000, 3000)(21000, 8000)
\put(15000, 5500){\vector(1,0){200}}
\qbezier(9000,8000)(15000,13000)(21000, 8000)
\put(15000,10500){\vector(1,0){200}}
\put(18000,13000){\circle{3000}}
\put(17600,14100){{\bf $\times$}}
\put(11700, 9600){{\bf $\times$}}
\put(14000,4000){\vector(1,0){2000}}
\put(16000,4000){$t$}
\put(9000,10000){$0$}
\put(21000,10000){$t$}
\put(12000, 8700){$t_1$}
\put(17600,13000){$t_2$}
\put(12000,11000){$J_\nu$}
\put(17800,15500){$J_\mu$}
\put(14500, 2000){(b)}
\put(23500,8800){$+~\bullet~\bullet~\bullet~\bullet$}
%
\end{picture}
\]
\caption{Quark skeleton diagrams similar to those in Fig. 1, except
that the two current insertions are on different quark lines. They
give higher twist contributions to $W_{\mu\nu}$.}
\end{figure}
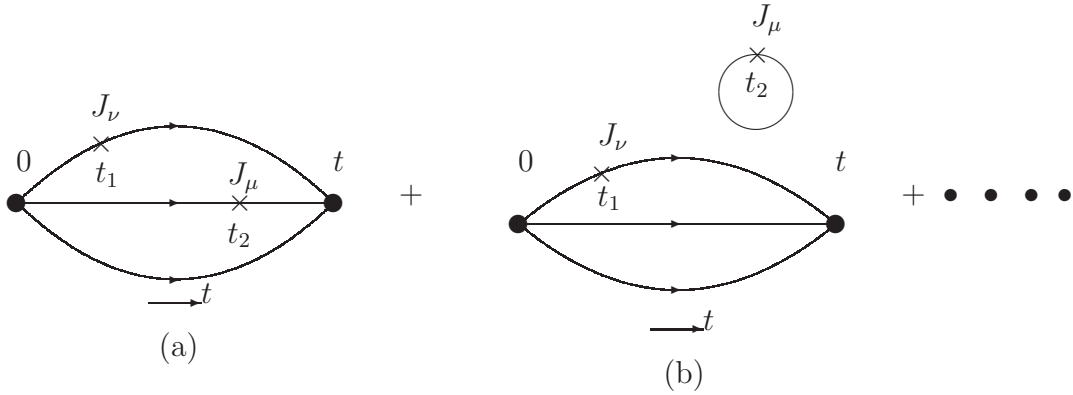

In the deep inelastic limit, the Bjorken scaling implies that
the current product (or commutator) is dominated by the light-cone 
singularity of a free-field theory, i. e. $1/x^2$ where  
$x^2 \approx O(1/Q^2)$. Among the time-fixed diagrams in
Fig.1, Fig. 1(a)/1(b) involves only a quark/antiquark propagator
between the currents; whereas, Fig. 1(c) has both quark and antiquark
propagators. Hence, there are two distinct classes of diagrams
where the sea quarks contribute. One comes 
from the D.I. ; the other comes from the C.I.. It is
usually assumed that connected insertions involve only ``valence"
quarks which are responsible for the baryon number. 
This is obviously not true, there are also quark-antiquark pairs in 
the C.I.  To define the quark distribution functions more
precisely, we shall call the quark/antiquark distribution from the
D.I. (which are connected to the ``valence''  quark propagators and
other quark loops through gluons) the ``disconnected sea'' quark/antiquark. 
We shall 
refer to the antiquark in the backward time going quark propagator between
$t_1$ and $t_2$ in Fig. 1(b) as the ``connected sea'' antiquark. On the
other hand, the quark in the time forward propagator between
$t_2$ and $t_1$ in Fig. 1(a) includes both the valence and the
``connected sea'' quarks. This is because a quark propagator from $t = 0$ to
$t = t (t > 0)$ involves both the time forward and backward zigzag motions so 
that one cannot tell if the quark propagator between $t_2$ and $t_1$ 
is due to the valence or the connected sea. All one knows is that it is a
quark  propagator. 
In other words, one needs to consider connected sea quarks in addition to the
valence in order to account for the production of quark-antiquark pairs
in  a connected fashion (Fig. 1(a)); whereas, the
pair production in a disconnected fashion is in Fig. 1(c).

We should stress that this separation into three topologically
distinct classes of path-integral diagrams is gauge invariant.
Notice that all the quark propagators are sewed together in a trace over
color.  
In a perturbative illustration of the distinction
between Fig. 1(b) and Fig. 1(c), one may consider the time-ordered
perturbation where Fig. 1(c) represents the vacuum 
polarization contribution as a disconnected insertion in a direct
diagram. The corresponding exchange
diagram where the quark in the loop in Fig. 1(c) is exchanged
with one in the ``valence'' will lead to a connected insertion
which falls in the class of Fig. 1(b)~\cite{ld94,jaf99}. 
However, the separation depends on the momentum
frame of the nucleon, although the sum which corresponds to the
full physical $W_{\mu\nu}(q^2,\nu)$ does not.
 For example, when the quark/antiquark
propagator between the currents is
either from the nucleon interpolation field or pair-produced
before the current $J_{\nu}$ at $t_1$, i.e. it is pre-existing
in the wavefunction, then it is not suppressed in the large 
momentum frame. Whereas, if it is pair-produced by the current
$J_{\nu}$, then it is suppressed by $|\vec{p}|^2$ where $|\vec{p}|$
is the momentum of the nucleon. This has been known
since current algebra sum rules were studied at large $|\vec{p}|$
~\cite{ad68}.   
 
Since the parton model acquires
its natural interpolation in the large momentum frame of the
nucleon, i.e. $|\vec{p}| \ge |\vec{q}|$, the parton distribution 
is then defined via $\nu W_2(Q^2, \nu) \longrightarrow F_2(x, Q^2) =
x \sum_i e_i^2 (q_i(x, Q^2) + \bar{q}_i(x, Q^2))$  in the large 
momentum frame. Here $x$ is the Bjorken scaling variable 
$x = Q^2/2m\nu$. Given the specific time-ordering in
Fig.1, Fig. 1(a)/1(b) involves only a quark/antiquark propagator
between the currents; whereas, Fig. 1(c) has both quark and antiquark
propagators. Consequently, the parton densities for the 
$u$ and $d$ antiquarks in the nucleon come from two sources, i.e. for the case
of $u$, 
\begin{equation}   \label{antiparton}
\overline{u}(x, Q^2) = \overline{u}_{cs}(x, Q^2) + \overline{u}_{ds}(x, Q^2),
\end{equation}
where $\overline{u}_{cs}(x, Q^2)$ is the ``connected sea'' (CS) u anti-parton
distribution from the  \mbox{C. I.} in Fig. 1(b) and $\overline{u}_{ds}(x,
Q^2)$ denotes the ``disconnected sea'' (DS) u anti-parton distribution from
the D. I. in Fig. 1(c). Similarly, $\bar{d}$ has two components. The strange
and charm partons, on the other hand, only appear in the D. I. in Fig. 1(c).
Thus, \begin{equation}  \label{sds}
\overline{s}(x, Q^2) \equiv \overline{s}_{ds}(x, Q^2).
\end{equation}
One can prove Eq. (\ref{sds}) this way. First, it is imperative to note that
the haronic tensor  $\widetilde{W}_{\mu\nu}(\vec{q}^{\,2},\tau)$ in Eq.
(\ref{wmunu_tilde}) does not  depend on the specific form of the interpolation
field, except its quantum numbers. In fact, the interpolation-field dependent
transition matrix element $f = \langle 0|O_N|N\rangle$ drops out in the ratio
of the four-point  to two-point functions in Eq. (\ref{wmunu_tilde}). As such,
one can use the simplest interpretation field of the nucleon which involves only
the valence quark field. For example, $O_N(t)$ can be taken to be the two $u$
and one $d$ quark fields with nucleon quantum numbers, 
\begin{equation}
O_N = \int d^3x e^{i \vec{p}
\cdot \vec{x}} \varepsilon^{abc} \Psi^{(u)a} (x) ((\Psi^{(u)b}(x))^TC
\gamma_5\Psi ^{(d)c}(x)), 
\end{equation}
for the proton. Since the interpolation field $O_N$ does not involve
strange quarks, the strange parton contribution can only come from the
vacuum polarization due to the external currents $J_{\mu}$ and $J_{\nu}$,
in other words, the D. I. in Fig. 1(c). As a corollary, one can also 
prove it for the case if one uses the $O_N s\bar{s}$ as the interpolation
field, for example. In this case, there are two classes of path-integral
diagrams. One class involves the \mbox{C. I.} where the strange quark
fields in the currents $\bar{s}\gamma_{\mu}s$ and $\bar{s}\gamma_{\nu}s$
contract with those strange quark fields in the interpolation field $O_N
s\bar{s}$ for the nucleon source and sink. This class of diagrams does not
project to the nucleon as its lowest mass state, since the physical states it
projects to will involve 5 valence quarks, i.e. $uuds\bar{s}$. Instead, it will
project to states such as the nucleon and a scalar $s\bar{s}$ meson. Since they
all have masses higher than the nucleon, they will be exponentially suppressed
relative to the nuelcon as the time separation $t - t_2$ and $t_1$ in Eq.
(\ref{wmunu_tilde}) are large. The other class involves a \mbox{D. I.} where
the strange quark fields in the nucleon source and sink self contract, so are
the strange quark fields in the currents. This will project to the nucleon
state with $uud$ as the valence quarks.  Since the transition matrix element
$f = \langle 0|O_N\bar{s}s|N\rangle$ is divided out in the ratio in Eq.
(\ref{wmunu_tilde}), it yields the same result as  that obtained with 
$O_N$ as the interpolation field. Thus, in either case, the strange parton
contribution comes only from the D. I. in Fig. 1 (c).

Similarly, the $u$ and $d$ partons have 2 sources, i.e.
\begin{equation}   \label{parton}
u(x, Q^2) = u_{v+cs}(x, Q^2) + u_{ds}(x, Q^2),
\end{equation}
where $u_{v+cs}(x, Q^2)$ denotes the valence and CS  $u$ partons and
$u_{ds}(x, Q^2)$ denotes the DS $u$ parton and they are from Fig. 1(a) and Fig.
1(c) respectively. Again these two components applies to the $d$ partons;
whereas, the $s$ parton has only the DS component.


\section{Operator product expansion (OPE)}

In the Minkowski space, the operator product expansion (OPE) is carried out 
in the unphysical region of $T_{\mu\nu}$ which is defined with 
the time-ordered product of the currents. How does one carry this out
in the Euclidean path-integral formulation? It turns out that because
$\widetilde{W}_{\mu\nu}(\vec{q}^{\,2},\tau)$ is defined in the 
Euclidean path-integral (Eq. (\ref{wtilde})), it requires several steps to
get to $T_{\mu\nu}$ in the Minkowski space. On the other hand, it is 
relatively 
easy to do so because it entails a simple Taylor expansion of functions 
as opposed to dealing with operators in the usual OPE, as we shall see. 
 
Considering Fig. 1(a) first, the three-point function in Eq.
(\ref{wmunu_tilde}) involves the following expression
\begin{equation}   \label{1a}
\widetilde{W}_{\mu\nu}(\vec{q}^{\,2},\tau) \propto \int d[A] det M(A) e^{-S_g}
Tr[ \cdots M^{-1}(t,t_2) \int d^3x e^{- i \vec{q}\cdot\vec{x}}\,i \gamma_{\mu}
M^{-1}(t_2, t_2 - \tau)\,i \gamma_{\nu} M^{-1} ( t_2 - \tau, 0) \cdots],
\end{equation}
where $S_g$ is the action for the gluon field $A$, $M^{-1}$ is the quark
propagator with arguments labeled by the Euclidean time. The spatial indices
are implicit and have been integrated over to give the nucleon a definite
momentum $|\vec{p}|$ and a momentum transfer $\vec{q}$. $\vec{x}$ and $\tau$
are the spatial and time  separations of the two currents $J_{\mu}$ and
$J_{\nu}$. The trace is over the color and spin indices. The expression
in Eq. (\ref{1a}) exhibits the part of the result from the quark line on which 
the currents are attached. The other two quark propagators and the nucleon
interpolation field operators are indicated by the dots.

 Similar to the usual OPE derivation~\cite{pes95}, we shall consider
the most singular part of the quark propagator between the currents in
Fig. 1. In the DIS limit where both the momentum transfer $|\vec{q}|$ and
energy transfer $\nu \rightarrow \infty$, the leading singularity
comes from the short-distance part in $\widetilde{W}_{\mu\nu}
(\vec{q}^{\,2},\tau)$ where $|\vec{x}|$ and $\tau \rightarrow 0$.
Therefore, we replace the quark propagator between the currents with
the free massless propagator
\begin{equation}
M^{-1}(t_2, t_2 - \tau) \longrightarrow \frac{1}{4 \pi^2} \frac{
\not{\partial}}{\vec{x}^2 + \tau^2}.
\end{equation}
We also carry out the Taylor expansion of the propagator $M^{-1}(
 t_2 - \tau, 0)$ for small $\tau$ and $\vec{x}$
\begin{equation}
M^{-1}(t_2 - \tau, 0) = e^{\vec{x}\cdot\vec{D} + \tau D_{\tau}}
M^{-1}(t_2, 0),
\end{equation}
where $D$ is the covariant derivative. With these substitutions, 
the corresponding hadronic tensor $W_{\mu\nu}(q^2, \nu)$ from Fig. 1(a)
after the Fourier transform in space and Laplace transform in $\tau$
(Eq. (\ref{wmunu})) is given as
\begin{equation}  \label{ft}
W_{\mu\nu}(q^2, \nu) \propto 
Tr[ \cdots M^{-1}(t,t_2)\,i \gamma_{\mu} \frac{-i \pi (\not{q} + i \not{D})}
{|\vec{q} + i \vec{D}|}
\delta (\nu + D_{\tau} - |\vec{q} + i \vec{D}|)\,i \gamma_{\nu}
M^{-1}(t_2, 0) \cdots].
\end{equation}

    Since $W_{\mu\nu}(q^2, \nu)$ is the imaginary part of $T_{\mu\nu}$, 
i. e. $W_{\mu\nu}(q^2, \nu) = \frac{1}{\pi} Im T_{\mu\nu}(q^2, \nu)$,
one can use the dispersion relation to obtain $ T_{\mu\nu}$ from
$ W_{\mu\nu}$,
\begin{eqnarray}  \label{tmn}
T_{\mu\nu}(q^2, \nu) = \frac{1}{\pi} \int_{Q^2/2 M_N + D_{\tau}}^{\infty}
d\nu' \frac{\nu' W_{\mu\nu}(q^2, \nu'- D_{\tau})}
{\nu'^2 - (\nu + D_{\tau})^2}, \nonumber \\
\propto Tr[ \cdots M^{-1}(t,t_2)\,i \gamma_{\mu} \frac{- i(\not{q} + i\not{D})}
{Q^2 + 2i\,q\cdot D - D^2}\,i \gamma_{\nu} M^{-1}(t_2, 0)\cdots],
\end{eqnarray} 
where we have used $\tau = i t$ and $D_t = i D_{\tau}$ so that
$D = (\vec{D}, -i D_t)$ is the covariant derivative in Minkowski space.
Expanding $T_{\mu\nu}$ in the unphysical region where
\mbox{$\frac{-2 p\cdot q}{Q^2} < 1$}, the expression between
the $\gamma's$ in Eq. (\ref{tmn}) gives
\begin{equation} \label{sde}
\frac{- i(\not{q} + i\not{D})}{Q^2 + 2i\,q\cdot D - D^2}
= \frac{- i (\not{q} + i\not{D})}{Q^2} \sum_{n =0}^{\infty} 
(\frac{- 2 i q\cdot D +  D^2}{Q^2})^n.
\end{equation}

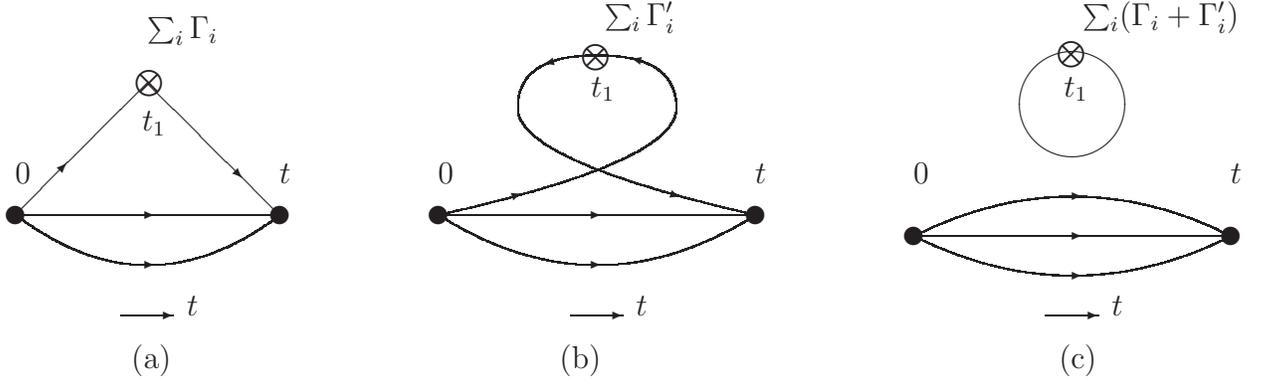
\begin{figure}[ht]
\[
\hspace*{1.5in}
\setlength{\unitlength}{0.01pt}
\begin{picture}(45000,20000)
\put(-10000,8800){\circle*{700}}
\put(-10000,8800){\line(1,1){5000}}
\put(-8232, 10568){\vector(1,1){200}}
\put(0000,8800){\line(-1,1){5000}}
\put(-1591,10391){\vector(1,-1){200}}
\put(-5500, 13500){{\bf $\bigotimes$}}
\put(-10000,8800){\line(1,0){10000}}
\put(-5000, 8800){\vector(1,0){300}}
\put(0000,8800){\circle*{700}}
\qbezier(-10000,8800)(-5000, 5000)(00,8800)
\put(-5000, 6900){\vector(1,0){300}}
\put(-6000, 5000){\vector(1,0){2000}}
\put(-5000,15500){{\bf $\sum_i \Gamma_i $}}
\put(-5200,12000){$t_1$}
\put(-3500, 5000){$t$}
\put(-10000,10000){$0$}
\put(0,10000){$t$}
\put(-5600, 3000){(a)}
\put(6000,8800){\circle*{700}}
\put(18000,8800){\circle*{700}}
\qbezier(9000, 12850)(9000 ,14850)(12000,14850)
\qbezier(12000,14850)(15000,14850)(15000,12850)
\qbezier(6000 ,8800)(15000,10800)(15000,12850)
\qbezier(9000 ,12850)(9000 ,10850)(18000,8800)
\put(11400,14500){{\bf $\bigotimes$}}
\put(13500,14700){\vector(-4,1){200}}
\put(10200,14560){\vector(-4,-1){200}}
\put(9000,  9550){\vector(3,1){200}}
\put(15000, 9550){\vector(3,-1){200}}
\put(12300,16000){{\bf $\sum_i \Gamma_i^\prime$}}
\put(6000,8800){\line(1,0){12000}}
\put(12000,8800){\vector(1,0){200}}
\qbezier(6000,8800)(12000, 5000)(18000,8800)
\put(12000, 6900){\vector(1,0){200}}
\put(11800,13300){$t_1$}
\put(13500, 5000){$t$}
\put(6000,10000){$0$}
\put(18000,10000){$t$}
\put(11000, 5000){\vector(1,0){2000}}
\put(10600, 3000){(b)}
\put(24000, 8000){\circle*{700}}
\put(24000, 8000){\line(1,0){12000}}
\put(30000, 8000){\vector(1,0){400}}
\put(36000, 8000){\circle*{700}}
\qbezier(24000, 8000)(30000, 5000)(36000, 8000)
\put(30000, 6500){\vector(1,0){400}}
\qbezier(24000, 8000)(30000,11000)(36000, 8000)
\put(30000, 9500){\vector(1,0){400}}
\put(30000,13000){\circle{4000}}
\put(29400,14600){{\bf $\bigotimes$}}
\put(30400,16000){{\bf $\sum_i (\Gamma_i+\Gamma_i^\prime)$}}
\put(29000, 5000){\vector(1,0){2000}}
\put(29700,13300){$t_1$}
\put(31500, 5000){$t$}
\put(24000,10000){$0$}
\put(36000,10000){$t$}
\put(29500, 3000){(c)}
%
\end{picture}
\]
\caption{Quark skeleton diagrams in the Euclidean path integral formalism
considered in the evaluation of matrix elements for the sum of local
operators from the
operator product expansion of $J_{\mu}(x) J_{\nu}(0)$. (a), (b) and (c)
corresponds to the operator product expansion from Fig. 1(a), 1(b) and
1(c) respectively.}
\end{figure}

  From this we obtain the valence and CS parton leading twist 
contributions to $T_{\mu\nu}$ from Fig. 1(a)
\begin{equation} \label{v+cs}
T_{\mu\nu}(q_{v+cs}) = \sum_f e_f^2 [8 p_{\mu} p_{\nu} \sum_{n =2}
\frac{(-2 q\cdot p)^{n -2}}{(Q^2)^{n -1}} A_f^n (C. I.) - 2 \delta_{\mu\nu}
 \sum_{n =2}\frac{(-2 q\cdot p)^n}{(Q^2)^n} A_f^n (C. I.)] + ...
\end{equation}
where $f$ indicates flavor. For the nucleon, it only involves $u$ and $d$ in
the C. I. $A_f^n (C. I.)$ is defined through the following consideration. We
first note that the short-distance expansion in Eq. (\ref{sde}) leads 
$T_{\mu\nu}(q_{v+cs})$ to a series of terms represented by the three-point
functions in \mbox{Fig. 3(a)} which correspond to matrix elements calculated
through the C. I. expression \begin{equation}
\int d[A] det M(A) e^{-S_g} Tr[\cdots M^{-1}(t, t_2) O_f^n M^{-1}(t_2,
0)\cdots], 
\end{equation}
where the operator $O_f^n$ is
\begin{equation}
O_f^n = i \gamma_{\mu 1} (\frac{-i}{2})^{n-1} \stackrel{\leftrightarrow}
{D}_{\mu 2} \stackrel{\leftrightarrow}{D}_{\mu 3} ...
\stackrel{\leftrightarrow}{D}_{\mu n}.
\end{equation}
The ratio of these three-point function in Fig. 3(a) to the
appropriate two-point functions then define the forward matrix
element which in turn defines the coefficient $A_f^n (C. I.)$
in Eq.(\ref{v+cs}),
\begin{equation}
\langle p|\bar{\Psi} O_f^n \Psi|p\rangle_{C. I.} = A_f^n (C. I.)
2 p_{\mu_1}p_{\mu_2} ... p_{\mu_n}.
\end{equation}

Similarly, we can perform the short-distance expansion for the CS antiparton 
in Fig. 1(b) and obtain the same expression as in Eq. (\ref{v+cs}) except with
the substitution $q \rightarrow - q$. As a result, this leads to the even n
terms minus the odd n terms instead of the sum as in Eq. (\ref{v+cs}), i. e.
\begin{equation}  \label{barcs}
T_{\mu\nu}(\bar{q}_{cs}) = \sum_{even = 2} \cdots A_f^n(C.I.) - 
\sum_{odd =3} \cdots A_f^n(C.I.).
\end{equation}   
In other words, the short-distance expansion of $T_{\mu\nu}$ from Fig. 1(b)
yields three-point functions with a series of insertion of the
same operators $\bar{\Psi} O_f^n \Psi$ except with a minus sign for the
odd $n$ terms. This is illustrated in Fig. 3(b) where $\Gamma_n'$ denotes
the n-th term insertion with the operator $O_f^n$ and the associated 
kinematic factors in Eq. (\ref{v+cs}). Comparing $\Gamma_n'$ in Fig. 3(b) 
and the corresponding $\Gamma_n$ in Fig. 3(a),
the minus sign for the odd $n$ terms
in Eq. (\ref{barcs}) implies that $\Gamma_n' = (-)^n \Gamma_n$.

By the same token, the short-distance expansion for the DS parton/antiparton 
contribution in $T_{\mu\nu}$ from Fig. 1(c) gives
\begin{equation} \label{ds/bards}
T_{\mu\nu}(q_{ds}/\bar{q}_{ds}) =  \sum_{even = 2} \cdots A_f^n(D.I.) \,\,\,+/-
\,\,\,\sum_{odd =3} \cdots A_f^n(D.I.).
\end{equation}
They have the same expression as $T_{\mu\nu}(q_{v+cs})$ and 
$T_{\mu\nu}(\bar{q}_{cs})$ except $A_f^n (D.I.)$ are from the D. I. part 
of the matrix element
\begin{equation}
\langle p|\bar{\Psi} O_f^n \Psi|p\rangle_{D. I.} = A_f^n (D. I.)
2 p_{\mu_1}p_{\mu_2} ... p_{\mu_n}.
\end{equation}
In this case, the leading twist expansion of the DS contribution to 
$T_{\mu\nu}$ in Fig. 1(c) now leads to two series of forward matrix
elements of D. I. One is for $T_{\mu\nu}(q_{ds})$ with even plus odd n 
terms; the other is  $T_{\mu\nu}(\bar{q}_{ds})$ with even minus odd n terms as
given in Eq. (\ref{ds/bards}). Both are represented in the three-point
functions in Fig. 3(c). It is worth pointing out that $T_{\mu\nu}(q_{v+cs})$
in Eq. (\ref{v+cs}) and $T_{\mu\nu}(q_{ds})$ in Eq. (\ref{ds/bards}) are
the same as those derived from the contraction of the inner pair of the
quark fields in the conventional operator product expansion of the
time-ordered current-current product 
$\bar{q}(x)\gamma_{\mu}q(x)\bar{q}(0)\gamma_{\nu}q(0)$~\cite{pes95}.
On the other hand, $T_{\mu\nu}(\bar{q}_{cs})$  
in Eq. (\ref{barcs}) and $T_{\mu\nu}(\bar{q}_{ds})$ in Eq. (\ref{ds/bards})
are the same as those from the contraction of the outer pair of the
quark fields in the current-current product. The only difference is that
the path-integral formalism allows the separation into the C.I. and the D.I.

When the parts in Eqs. (\ref{v+cs}), (\ref{barcs}), and (\ref{ds/bards}) are
summed up, only the even $n$ terms of the OPE are left
\begin{eqnarray}  \label{sumtmu}
T_{\mu\nu}& =& T_{\mu\nu}(q_{v+cs}) + T_{\mu\nu}(\bar{q}_{cs}) 
+ T_{\mu\nu}(q_{ds} + \bar{q}_{ds})   \nonumber \\
 &=& 2 \sum_{n=2, even} \cdots (A_f^n (C. I.) + A_f^n (D. I.)).
\end{eqnarray}
This is the same as derived from the ordinary OPE. However, 
what is achieved with the path-integral formulation is the separation 
of C. I. from D. I. in addition to the separation of 
partons from antipartons. This has not been possible with other
formulations, e.g. the light-cone definition of the distribution function.
This separation facilitates the derivation of the different small x
behavior between the CS and the DS, the identification of
the CS parton as the source of the Gottfried sum rule violation, and a
different evolution of $\bar{q}_{cs}(x, Q^2)$ from  that of $q_{ds}(x, Q^2)$
and $\bar{q}_{ds}(x, Q^2)$ as we shall see.

    Now consider the contour integral of $T_2$ in the C. I. and D. I. parts
of $T_{\mu\nu}$ around $\nu = 0$ in the complex $\nu$ plane while keeping
$Q^2$ fixed. $\oint \frac{d \nu}{2 \pi i} \frac{T_2 (\nu, Q^2)}{\nu^{n -1}}$
picks up the $\nu^{n -2}$ term in the series expansion of $T_2$ in
Eqs. (\ref{v+cs}), (\ref{barcs}), and (\ref{ds/bards}) (N.B. $- 2 q\cdot p
= 2 M_p \nu$ in these equations.)
\begin{equation} \label{ci1}
\oint \frac{d \nu 2 M_p}{2 \pi i} \frac{T_2 (\nu, Q^2)}{\nu^{n -1}}
= \sum_f 8 e_f^2 (\frac{2 M_p}{Q^2})^{n -1} A_f^n,
\end{equation}
for both the C. I. and the D. I. The contour of the integral can be
distorted to turn the above integral into an integral over the 
discontinuities of $T_2$. Through the dispersion relation, this gives
\begin{equation}   \label{ci2}
2 \int_{Q^2}^{\infty} \frac{d \nu 2 M_p}{2 \pi i} \frac{ 2 i 
\,Im\, T_2 (\nu, Q^2)}{\nu^{n -1}} = 8 (\frac{2 M_p}{Q^2})^{n -1}
\int_0^1 dx x^{n-2} \frac{2 M_p \nu W_2(\nu, Q^2)}{4 }.
\end{equation}

    Equating these two integrals and relating $W_2$ to the 
parton distribution function, one obtains the moment sum rules.
Since $T_2(Q^2,\nu)$ is symmetric w.r.t $\nu \rightarrow
-\nu$, we obtain only the sum rules for even $n$.
Thus,
\begin{eqnarray}    \label{even}
A_f^{n = even} (C. I.) = M^n (C. I.)\equiv  \int_0^1 dx\, x^{n -1} 
(q_{v+cs}(x, Q^2) + \bar{q}_{cs} (x, Q^2)),      \nonumber \\         
A_f^{n = even} (D. I.) = M^n (D. I.)\equiv \int_0^1 dx\,  x^{n -1} 
(q_{ds}(x, Q^2) + \bar{q}_{ds}(x, Q^2)).
\end{eqnarray}
Similarly we obtain the moment sum rules for the odd $n$ from the $W_3$ form
factor through the interference of the vector and axial vector currents,
\begin{eqnarray}  \label{odd}
A_f^{n = odd} (C. I.) = M^n (C. I.)\equiv\int_0^1 dx\, x^{n -1} 
(q_{v+cs}(x, Q^2) - \bar{q}_{cs} (x, Q^2)),      \nonumber \\
A_f^{n = odd} (D. I.) = M^n (D. I.)\equiv \int_0^1 dx\, x^{n -1} 
(q_{ds}(x, Q^2) - \bar{q}_{ds}(x, Q^2)).
\end{eqnarray}

One can define the valence parton distribution
\begin{equation} \label{val_def}
q_v (x, Q^2) \equiv q_{v+cs} (x, Q^2) - \bar{q}_{cs} (x, Q^2).
\end{equation}
In this case, $A_f^{n = odd}$ gives the sum rules for the
valence distribution. In particular, valence number sum rules 
\begin{eqnarray}
M_u^1 (C. I.) \equiv  \int_0^1 dx\, u_v (x, Q^2) = 2, \nonumber \\
M_d^1 (C. I.) \equiv  \int_0^1 dx\, d_v (x, Q^2) = 1, \nonumber \\ 
M_f^1 (D. I.) \equiv  \int_0^1 dx\, (q_{ds}(x, Q^2) - \bar{q}_{ds}(x, Q^2))
  = 0,
\end{eqnarray}
for the $u$ and $d$ quarks in the proton 
reflect the charge conservation of the vector current
$\Psi\,i \gamma_{\mu}\Psi$ and the fact that the DS carries no net charge.

     We note that the matrix elements associated with $A_f^{n = even}
(C. I.)$ include not just the valence but also the CS contribution.
This is why the matrix element $A_{u,d}^2 (C. I.)$, which corresponds to
the momentum fraction $M^2 (C. I.) = \langle x\rangle_{C. I.}$, when calculated
on the lattice ~\cite{ms89,goe96}, is larger than that obtained from the
experiments  for the valence partons only,  \mbox{i.e.} $\langle x\rangle_{C.
I.} > \langle x\rangle_v$ at $\mu \sim 2$ GeV.

\subsection{Operator Rescaling and Mixing, and Parton Evolution}

     The dimensionless coefficients $A_f^n$ are not constants, but
depend logarithmically on $Q^2$, the renormalization point of the operator 
product expansion. The operator rescaling and mixing analysis~\cite
{gp74,gw74} for the twist-two flavor non-singlet and singlet operators 
gives the renormalization group equations for the corresponding moments of
the structure functions. In the context of the present path-integral 
formulation of OPE, the non-singlet coefficients
$A_f^n$ and the non-singlet moments only have contributions from the C. I.   
(Fig. 3(a) or Fig. 3(b)), since their D. I. contributions cancel among 
different flavors. On the other hand, the singlet coefficients $A_f^n$ and
the singlet moments have both the C. I. and D. I. (Fig. 3(c)) contributions.
Therefore, it is possible to go to the single flavor basis and classify
the equations in terms of C. I. and D. I. For the C. I. which involves
only the valence flavors (e.g. $u$ and $d$ for the nucleon), the
renormalization equation is
\begin{equation}  \label{mci}
\frac{d}{d\,\ln Q^2} M_f^n (C. I.) = \frac{\alpha_s(Q^2)}{8 \pi}
a_{qq}^n M_f^n (C. I.),
\end{equation}
where $a_{qq}^n$ is the anomalous dimension coefficient.
For the D. I. the equation is
\begin{equation}  \label{mdi}
\frac{d}{d\,\ln Q^2} M_f^n (D. I.) = \frac{\alpha_s(Q^2)}{8 \pi}
(a_{qq}^n M_f^n (D. I.) + \frac{1 + (-)^n}{2} a_{qG}^n M_G^n),
\end{equation}
where $ a_{qG}^n$ is the anomalous dimension coefficient for operator mixing
with the gluon operators and $M_G^n$ is the moment for the gluon distribution
function. We note that this mixing with gluon operators only contributes
to $n = even$. As we can see, \mbox{Eq. (\ref{mdi})} involves the DS parton
only.

We should stress that in the literature~\cite{ap77,cq89,ehl84}
the non-singlet case has frequently been identified with valence. This is
clearly incorrect. As we see from \mbox{Eq. (\ref{even})} that $A_f^{n =
even}(C. I.)$ includes the CS partons in addition to the valence partons.
Detailed study of this subject on the lattice has been carried out for the
matrix elements  and form factors of the nucleon from the three-point
functions as well as hadron masses from the two-point functions~\cite{ldd99}.
It is shown when the  CS quarks are removed by prohibiting pair-production
through the Z  graphs in the C. I., the hadron structure and masses are
greatly affected.  It is learned that the CS quarks are responsible for the
meson dominance in the form factors, the deviation of $F_A/D_A$ and $ F_S/D_S$
from the  non-relativistic $SU(6)$ limit, the hyper-fine splittings, and the
constituent quark masses~\cite{ldd99}. In the context of the parton model, they
are responsible for the difference between $\bar{u}(x)$ and $\bar{d}(x)$ in the
proton.

   Following Altarelli and Parisi~\cite{ap77}, the rescaling and mixing equations 
in Eqs. (\ref{mci}) and (\ref{mdi}) can be translated into integral-differential
equations which are the evolution equations for the parton densities.
Therefore, for the C. I. the evolution equation for the unpolarized valence
and CS partons is  
\begin{equation}  \label{evcs}
\frac{d\,q_{v+cs}(x, Q^2)}{d\,\ln Q^2} = \frac{\alpha_s (Q^2)}{2 \pi}
\int_x^1 \frac{d\,y}{y} P_{qq}(\frac{x}{y}) q_{v+cs}(y, Q^2),
\end{equation}
where 
\begin{equation}   \label{adc}
\int_0^1 d\,z z^{n-1} P_{qq} (z) = \frac{a_{qq}^n}{4}.   
\end{equation}
For the CS antiparton density, the equation is similar
\begin{equation}    \label{ecs}
\frac{d\,\bar{q}_{cs}(x, Q^2)}{d\,\ln Q^2} = \frac{\alpha_s (Q^2)}{2 \pi}
\int_x^1 \frac{d\,y}{y} P_{qq}(\frac{x}{y}) \bar{q}_{cs}(y, Q^2).
\end{equation}
For the DS partons in the D. I. (Fig. 1(c)), the evolution equations 
from Eq. (\ref{mdi}) are
\begin{equation} \label{eds}
\frac{d\,(q_{ds} + \bar{q}_{ds})(x, Q^2)}{d\,\ln Q^2} = \frac{\alpha_s 
(Q^2)}{2 \pi}
\int_x^1 \frac{d\,y}{y} [P_{qq}(\frac{x}{y}) (q_{ds} + \bar{q}_{ds})(y, Q^2) +
P_{qG} (\frac{x}{y}) G (y, Q^2)],
\end{equation}
\begin{equation}
\frac{d\,(q_{ds} - \bar{q}_{ds})(x, Q^2)}{d\,\ln Q^2} = \frac{\alpha_s 
(Q^2)}{2 \pi}
\int_x^1 \frac{d\,y}{y} P_{qq}(\frac{x}{y}) (q_{ds} - \bar{q}_{ds})(y, Q^2),
\end{equation}
where 
\begin{equation}
\int_0^1 d\,z z^{n-1} P_{qG} (z) = \frac{a_{qG}^n}{4},
\end{equation}
and $G(y, Q^2)$ is the unpolarized gluon distribution function.  

Finally, the gluon evolution equation is
\begin{eqnarray}
\frac{d\,G(x, Q^2)}{d\,\ln Q^2} & =& \frac{\alpha_s (Q^2)}{2 \pi}
\int_x^1 \frac{d\,y}{y} \{P_{Gq}(\frac{x}{y}) [\sum_{f = val. fla.}
(q_{v+cs}^f + \bar{q}_{cs}^f)(y, Q^2) + \sum_{f = DS fla.}(q_{ds}^f +
\bar{q}_{ds}^f)(y, Q^2)]  \nonumber \\
&+& P_{GG} (\frac{x}{y}) G (y, Q^2)\},
\end{eqnarray}

    It appears that, except for the gluon distribution, the evolution equations
derived above are different from the DGLAP evolution equations {\it a la}
Dokshitzer~\cite{dok77}, Gribov and Lipatov~\cite{gl72}, and Altarelli and
Parisi~\cite{ap77} due to an extra CS degree of freedom in Eq.(\ref{ecs}) which
evolves like the valence in Eq. (\ref{evcs}). However, 
since the evolution equations are linear in the parton distribution functions,
once one defines the total sea to be the sum of CS and DS, i.e.
\begin{eqnarray}  
q_s(x, Q^2) \equiv q_{cs}(x, Q^2) + q_{ds}(x, Q^2), \nonumber \\
\bar{q}_s(x, Q^2) \equiv \bar{q}_{cs}(x, Q^2) +\bar{q}_{ds}(x, Q^2),\nonumber \\ 
\end{eqnarray}
 the sum of twice of Eq. (\ref{ecs}) and Eq. (\ref{eds})
becomes 
\begin{equation}  \label{ets}
\frac{d\,(q_s + \bar{q}_s)(x, Q^2)}{d\,\ln Q^2} = \frac{\alpha_s 
(Q^2)}{2 \pi}
\int_x^1 \frac{d\,y}{y} [P_{qq}(\frac{x}{y}) (q_s + \bar{q}_s)(y, Q^2) +
P_{qG} (\frac{x}{y}) G (y, Q^2)].
\end{equation}
This is exactly the evolution of sea partons in the DGLAP equation.
Similarly, the difference between Eq. (\ref{evcs}) and Eq. (\ref{ecs}) leads to
the evolution of the valence partons in the DGLAP equation,
\begin{equation}   \label{eval}
\frac{d\,q_v(x, Q^2)}{d\,\ln Q^2} = \frac{\alpha_s (Q^2)}{2 \pi}
\int_x^1 \frac{d\,y}{y} P_{qq}(\frac{x}{y}) q_v(y, Q^2).
\end{equation}

    Thus, we see DGLAP equations can be derived from the path-integral
formalism. However, the present form with the separation of CS from 
the DS offers a separate evolution of the CS in Eq. (\ref{ecs}) which
are decoupled from the valence, the DS, and the gluon. This decoupling
is retained when higher orders are considered~\cite{fp82}. This 
affords the possibility of maintaining the separation of the CS and DS
parton distributions when they are fitted to experiments at certain $Q_0^2$ and
evolved up or down in $Q^2$. This
is consistent with the classification into valence, CS, and DS in Fig. 1 and
Eqs. (\ref{antiparton}) and (\ref{parton}) at all $Q^2$ from the path-integral
formalism in the first place. Although, at leading twists, it makes no
difference whether one considers the separate evolutions of CS and DS in
Eqs. (\ref{ecs}) and (\ref{eds}) or the combined evolution in Eq. (\ref{ets}),
it may be necessary to consider the separate evolutions of CS and DS when 
higher twists are taken into account~\cite{guoqiu}.

\section{Phenomenological Consequences}

    After the dynamical parton degrees of freedom are classified 
as valence, connected sea (CS) , and disconnected sea (DS) via the
path-integral diagrams in Fig. 1, one may question if there is an advantage to
separating the CS from the DS. After all, both the CS and DS are pair-produced
sea, only with different topology in the path-integral diagrams. 
In the previous section, we have shown that the CS parton evolves the
same way as the valence, i.e. it is decoupled from the DS and the gluons in the 
evolution equation. This is consistent with the classification of the
parton degrees of freedom into valence, CS, and DS at all $Q^2$.
In the following, we shall show
that it is useful to distinguish CS from DS because they have
different small x behaviors. This is especially important in view of
the fact that, in the nucleon, $\bar{u}$ and $\bar{d}$ have both
the CS and DS parts, yet $\bar{s}$ has only the DS part.
We further note that the Gottfried sum rule violation is attributable to 
the CS partons primarily. We finally asses the magnitude of the momentum
fraction due to the CS and DS partons.

\subsection{Small x behavior}
 
Since the CS parton is in the connected insertion which is flavor non-singlet
like the valence, its small x behavior reflects the leading Reggeon exchanges
of $\rho, \omega, a_2...$ and thus should be like
$x^{-1/2}$~\cite{kw71,mof72,rey81,bs91}. On the other hand, the DS is flavor singlet
and can have Pomeron exchanges, its small x behavior goes like
$x^{-1}$~\cite{kw71,mof72,rey81}.  Therefore,  we have 
\begin{eqnarray}
\bar{u}_{cs}/\bar{d}_{cs}(x, Q_0^2)_{\stackrel{\sim}{x \rightarrow 0}} 
x^{-1/2}, \\
 \bar{u}_{ds}(x, Q_0^2) \sim \bar{d}_{ds}(x, Q_0^2) \sim \bar{s}(x, Q_0^2)_
{\stackrel{\sim}{x \rightarrow 0}} x^{-1}.
\end{eqnarray}
at certain $Q_0$. As a result, 
\begin{equation}
\bar{u}(x, Q_0^2) - \bar{d}(x, Q_0^2)_ {\stackrel{\sim}{x \rightarrow 0}} x^{-1/2},
\end{equation}
and this has been incorporated in the fitting of experimental results at
some input scale $Q_0^2$~\cite{cteq4,cteq99,mrs98}. On the other hand, it has been 
taken for granted that  $\bar{u}(x, Q_0^2) + \bar{d}(x, Q_0^2)$ has the
same behavior as $\bar{s}(x, Q_0^2)$ in the fitting of the experiments.
In other words, it is assumed that
\begin{equation}
\frac{1}{2}(\bar{u}(x, Q_0^2) + \bar{d}(x, Q_0^2)) \sim 2 \bar{s}(x, Q_0^2),
\end{equation}
in the fitting of parton distributions at  $Q_0^2 =
1.6 \,{\rm GeV^2}$ for CTEQ4~\cite{cteq4} and $Q_0^2 = 1 \,{\rm GeV^2}$ for
MRST~\cite{mrs98}. This has made the assumption that both the $\bar{u},
\bar{d}$, like $\bar{s}$, have only the DS component. We proved in Section
\ref{pif} (Eqs. (\ref{antiparton}) and (\ref{parton})) that this is not true
and that $\bar{u}$ and $\bar{d}$ have CS in addition to DS. Therefore, the
correct parametrization for  $\frac{1}{2}(\bar{u}(x, Q_0^2) + \bar{d}(x,
Q_0^2))$ should be 
\begin{equation} \label{udsum} 
\frac{1}{2}(\bar{u}(x, Q_0^2)
+ \bar{d}(x, Q_0^2)) = A \bar{s}(x, Q_0^2)  + \frac{1}{2}(\bar{u}_{cs}(x,
Q_0^2) + \bar{d}_{cs}(x, Q_0^2)), \end{equation}
where $A$ is a proportional constant. In the $SU(3)$ flavor symmetric limit,
$ A = 1$.The second term on the right hand side of Eq. (\ref{udsum})
is the CS part and has the $x^{-1/2}$ small x behavior. As far as we
know, this kind of parametrization for $\frac{1}{2}(\bar{u}(x, Q_0^2) +
\bar{d}(x, Q_0^2))$ has not been taken into account in extracting the parton
distribution functions. When this CS degree of freedom is incorporated in the
sum of $\bar{u}$ and $\bar{d}$ in addition to the difference, we will have a
different result from the present global analyses which will lead to
different predictions on the parton distributions at high $Q^2$ relevant
to LHC.

\subsection{Origin of Gottfried sum rule violation}

   The NMC experiments of the $F_2$ structure functions of the proton
and deuteron~\cite{nmc92} reveal that the Gottfried sum rule
\begin{equation}
S_G =  \int_0^1 dx \frac{F_2^p(x) - F_2^n(x)}{x} = \frac{1}{3},
\end{equation}
is violated due to fact that $\bar{u} \neq \bar{d}$ in the proton. This
is verified in the Drell-Yan experiment E866/NuSea~\cite{pen98}.   
 It has been shown~\cite{ld94} that 
the DS partons in Fig. 1(c) cannot give rise to a different $\bar{u}$ and
$\bar{d}$ when $m_u = m_d$. It is noted that in Fig. 1(c) the flavor indices in
the quark loop are separately traced from those in the propagators associated
with the nucleon interpolation field, only the latter reflects the valence
nature of the proton. Hence, Fig. 1(c) does not distinguish a loop with an u
quark  from that with a d quark at the flavor symmetric limit, i.e.
$m_u = m_d$. In other words, $\bar{u}_{ds}(x, Q^2) = \bar{d}_{ds}(x, Q^2)$. 
On the other hand, the origin of $\bar{u}(x, Q^2) \neq
\bar{d}(x, Q^2)$ can come from the CS antipartons in Fig. 1(b).
Thus, the violation of the Gottfried sum rule originates entirely from
the CS partons in the charge symmetric limit, i.e.
\begin{equation}  \label{gsr}
 \int_0^1 dx \frac{F_2^p(x, Q^2) - F_2^n(x, Q^2)}{x} = \frac{1}{3} +
\frac{2}{3}\int dx [\bar{u}_{cs}(x, Q^2) - \bar{d}_{cs}(x, Q^2)].
\end{equation}       
We shall see later that to $O(\alpha_s)$ and $ O(1/Q^2)$, this turns 
out to be a sum rule.

 Due to the fact that 
\begin{equation}
\int_0^1 dx P_{qq}(x) = \frac{a_{qq}^1}{4} = 0,
\end{equation}
where $a_{qq}^1$ is the anomalous dimension coefficient in Eq. (\ref{adc}),
Eqs. (\ref{eval}) and (\ref{ecs}) lead to 
\begin{eqnarray}
\frac{d}{d \ln Q^2} \int_0^1 dx\, q_v (x, Q^2) = 0, \label{valc} \\
\frac{d}{d \ln Q^2} \int_0^1 dx\, \bar{q}_{cs} (x, Q^2) = 0.  \label{cloudc}
\end{eqnarray}
We see that the CS antiparton number, like the valence number, is 
conserved, i.e. independent of $Q^2$. This is in contrast with the DS
partons whose number is not conserved due to the pair-creation from the
gluon.  However, the conservation of the CS antiparton number is only good in
the leading logarithmic approximation, i.e. good to $O(\alpha_s)$ and 
$O(1/Q^2)$. Whereas, the conservation of the vector current protects the
charges, thus the valence quark number, against any $Q^2$ 
correction~\cite {gp74,gw74,ap77}.

    We note that the sum in Eq. (\ref{gsr}) is in terms of the CS 
antipartons numbers. Thus to leading logarithmic approximation, it is a
sum rule
\begin{equation}   \label{mysr}
 \int_0^1 dx \frac{F_2^p(x) - F_2^n(x)}{x} = \frac{1}{3} +
\frac{2}{3} [n_{\bar{u}_{cs}} - n_{\bar{d}_{cs}}],
\end{equation}
where $n_{\bar{u}_{cs}}/n_{\bar{d}_{cs}}$ is the $\bar{u}_{cs}/\bar{d}_{cs}$
number.

\subsection{Magnitude of the connected sea partons}

      We don't know precisely how large the magnitude of the CS partons and
antipartons are in comparison with those of the valence and the DS unless one
fits the DIS and Drell-Yan experiments with an explicit separation of the CS
and the DS. But there are hints which are helpful in this respect. 
From the NMC experiments on the $F_2$ structure function 
of the proton and deuteron~\cite{nmc92}, the sum in Eq. (\ref{mysr})
is measured to be $0.235 \pm 0.026$ at $Q^2 = 4 \, {\rm GeV^2}$, significantly
smaller than the Gottfried sum rule prediction of 0.333. This implies that
\begin{equation}
n_{\bar{u}_{cs}} - n_{\bar{d}_{cs}} = \int_0^1 dx [\bar{u}_{cs}(x, Q^2) -
\bar{d}_{cs}(x, Q^2)]  = - 0.147 \pm 0.039,
\end{equation}
at $Q^2 = 4 \,{\rm GeV^2}$ which is not negligible compared with the valence
numbers of $u$ and $d$.

    Since $\bar{u}$ (similarly $\bar{d}$) has contributions from the 
CS and DS, i. e. \mbox{$\bar{u}(x) = \bar{u}_{cs}(x) + \bar{u}_{ds}(x)$} and
$\bar{s}$ is from the DS only, one expects that
\begin{equation} 
\langle x\rangle_{\bar{u}} = \langle x\rangle_{\bar{u}_{cs}} +
\langle x\rangle_{\bar{u}_{ds}} > \langle x\rangle_{\bar{s}}. 
\end{equation}
Indeed, in the CCFR dimuon data, $\bar{s}$ is about 50\% of 
$(\bar{u} + \bar{d})/2$ at $Q^2 \simeq 4\, {\rm GeV^2}$ in the range of 
$x$ between 0.01 and 0.20. Since $\bar{u} - \bar{d}$, which only has
the CS contribution, also peaks in this x range, we think the 
observed difference between $(\bar{u} + \bar{d})/2$ and $\bar{s}$ is
mainly duw to the CS part of $(\bar{u} + \bar{d})/2$. As a result, one
expects that the momentum fraction of the CS part of
$(\bar{u} + \bar{d})/2$ is comparable to that of the DS part, i.e.
\begin{equation}
\frac{1}{2}(\langle x\rangle_{\bar{u}_{cs}}+ \langle x\rangle_{\bar{d}_{cs}})
 \sim \frac{1}{2}(\langle x\rangle_{\bar{u}_{ds}}+ \langle
x\rangle_{\bar{d}_{ds}}) \sim \langle x\rangle_{\bar{s}},
\end{equation}
at $Q^2 = 4\, {\rm GeV^2}$. Any experiment which measures
$(\bar{u} + \bar{d})/2$ and $\bar{s}$ at very small x, e.g. $10^{-3} -
10^{-4}$ will be very useful in verifying the form of the distribution
function prescribed in Eq. (\ref{udsum}).

\section{Conclusion}  
            
In conclusion, we have formulated the hadronic tensor $W_{\mu\nu}$ of the
deep inelastic scattering starting from the Euclidean path-integral
formalism. We found that it can be divided into three gauge-invariant and
topologically distinct parts which we classify as the valence-connected sea
partons, the connected sea antipartons and the disconnected-sea partons and
antipartons. This admits a separation of the C. I. from the D. I. and the
partons from the antipartons. Since the CS is in the  C. I. and the DS in
the D. I., they have different small x behaviors. We show that the operator
product expansion is simply a short distance Taylor expansion of functions
in the path-integral. From operator rescaling and mixing, we derive
the evolution equations which show that the CS partons evolve like the
valence and their numbers are conserved in the leading log approximation.
We stress 
that in the nucleon  $\bar{u}$ and $\bar{d}$ partons have both the connected
and disconnected sea contributions; whereas, $\bar{s}$ partons has only the
disconnected sea contribution. A global
analysis of the experimental data is needed to take this difference into
account to fit the parton distribution functions.

This work is partially supported by U.S. DOE grant No. DE-FG05-84ER40154. 
The author would like to thank S. Brodsky, N. Christ, G. T. Garvey,
X. Guo, \mbox{X. Ji}, \mbox{ L. Mankiewicz,} R. McKeown,
C.S. Lam, G. Martinelli, J.C. Peng, J. Qiu, \mbox{W.K. Tung} and C.P. Yuan for
useful discussions.


\begin{thebibliography}{99}
\bibitem{emc88}
J. Ashman et al. (EMC), Phys. Lett. {\bf B206}, 364 (1988);
K. Abe et al. (E143),  Phys. Rev. Lett. {\bf 74}, 346 (1995);
D. Adams et al. (SMC), Phys. Rev. {\bf D 56}, 5330 (1997).
\bibitem{nmc92}
NMC Collaboration, P. Amaudrux et al., Phys. Rev. Lett {\bf 66}, 2717 (1992);
M. Arneodo et al., Phys. Rev. {\bf D50}, R1 (1994).
\bibitem{ld94}
K.F. Liu and S.J. Dong, Phys. Rev. Lett., {\bf 72}, 1790 (1994).
\bibitem{ldd99}
K. F. Liu, S. J. Dong, T. Draper, D. Leinweber, J. Sloan, W. Wilcox,
and R. M. Woloshyn, Phys. Rev. {\bf D59}, 112001 (1999).
\bibitem{wil93}
W. Wilcox, Nucl. Phys. (Proc. Suppl.), {\bf B30}, 491 (1993).
\bibitem{jaf99}
This point was brought up by R. Jaffe during the Gordon Conference
on Nuclear Physics and QCD in July 1999.
\bibitem{ad68}
See, for example, S. Adler and R. Dashen, {\it Current Algebra and 
Applications to Particle Physics}, pp. 254 (Benjamin, N. Y., 1968).
\bibitem{pes95}
We shall follow the derivation of operator analysis in
M. E. Peskin and D. V. Schroeder, \underline{Quantum Field Theory},
Addison-Weslay, 1995.
\bibitem{ms89}
G. Martinelli and C. T. Sachrajda, Nucl. Phys. {\bf B316}, 355 (1989).
\bibitem{goe96}
M. G\"{o}ckeler, et al., Phys. Rev. {\bf D53}, 2317 (1996).
\bibitem{gp74}
H. Georgi and H. D. Politzer, Phys. Rev. {\bf D 9}, 416 (1974).
\bibitem{gw74}
D. Gross and F. Wilczek, Phys. Rev. {\bf D 9}, 980 (1974).
\bibitem{ap77}
G. Altarelli and G. Parisi, Nucl. Phys. {\bf B126}, 298 (1977).
\bibitem{dok77}
Yu. Dokshitzer, Sov. Phys. JETP {\bf 46}, 641 (1977).
\bibitem{gl72}
V.N. Gribov and L.N. Lipatov, Sov. J. Nucl. Phys. {\bf 15}, 438, 675 (1972).
\bibitem{cq89}
J. C. Collins and J. Qiu, Phys. Rev. {\bf D39}, 1398 (1989).
\bibitem{ehl84}
E. Eichten, et al., Rev. Mod. Phys. {\bf 56}, 579 (1984).
\bibitem{fp82}
W. Furmanski and R. Petronzio, Z. Phys. {\bf C11}, 293 (1982).
\bibitem{guoqiu}
X. Guo and J. Qiu, private communication.
\bibitem{kw71}
J. Kuti and V.F. Weissopf, Phys. Rev. {\bf D 4}, 3418 (1971).
\bibitem{mof72}
J.W. Moffat, Schladming lectures 1972, Acta Physica Austriaca, Suppl. IX, 
605 (1972).
\bibitem{rey81}
E. Reya, Phys. Rep. {\bf 69}, 195 (1981).
\bibitem{bs91}
S. Brodsky and I. Schmidt, Phys. Rev. {\bf D43}, 179 (1991).
\bibitem{cteq4}
H. L. Lai, et al., Phys. Rev. {\bf D55}, 1280 (1997). 
\bibitem{cteq99}
H. L. Lai, et al., hep-ph/9903282.
\bibitem{mrs98}
A.D. Martin, R.G. Roberts, W.J. Stirling, and R.S. Thorne,
Eur.Phys.J.{\bf C4}, 463 (1998). 
\bibitem{pen98}
E866/NuSea Collaboration, J. C. Peng et al., Phys. Rev. {\bf D58}, 092004 
(1998).
\end{thebibliography}
\end{document}